\documentclass[prb,twocolumn]{revtex4}

\usepackage{amsmath,amssymb,graphicx}

\begin{document}

\title{Bit storage by $360^\circ$ domain walls in ferromagnetic
  nanorings} 

\author{C. B. Muratov}

\affiliation{Department of Mathematical Sciences,
    New Jersey Institute of Technology, Newark, NJ 07102} 

\author{V. V. Osipov}

\affiliation{Mission Critical Technologies, Inc., 2041
    Rosecrans Avenue, Suite 225, El Segundo, CA 90245 }

\affiliation{Intelligent Systems Division, D\&SH Branch, NASA Ames
    Research Center, MS 269-1, Moffett Field, CA 94035}

\date{\today}

\begin{abstract}
  We propose a design for the magnetic memory cell which allows an
  efficient storage, recording, and readout of information on the
  basis of thin film ferromagnetic nanorings. The information bit is
  represented by the polarity of a stable 360$^\circ$ domain wall
  introduced into the ring. Switching between the two magnetization
  states is achieved by the current applied to a wire passing through
  the ring, whereby the $360^\circ$ domain wall splits into two
  charged $180^\circ$ walls, which then move to the opposite extreme
  of the ring to recombine into a $360^\circ$ wall of the opposite
  polarity.
\end{abstract}


\maketitle

Magnetoresistive Random Access Memory (MRAM) has long been considered
as one of the main contenders to replace the current
semiconductor-based devices as a universal, non-volatile computer
memory \cite{akerman05,dennis02,tehrani03,hubert}. Nevertheless,
despite recent developments of commercially available products
\cite{akerman05}, MRAM technology is yet to deliver a device that
would truly compete with the existing semiconductor technologies. 

\begin{figure}
  \centering
  \includegraphics[width=3.1in]{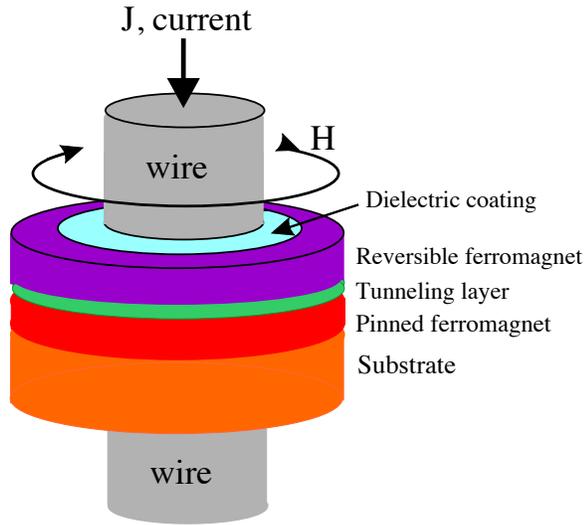}
  \caption{Schematics of the proposed design.}
  \label{fig:design}
\end{figure}

In the current MRAM cell designs the storage element is a sandwich
structure which contains two ferromagnetic layers separated by a thin
tunneling layer; the magnetization is fixed in one of the
ferromagnetic layers, but is free to move in the
other\cite{tehrani03,akerman05}. The bit of information is encoded as
a monodomain magnetization state of the free ferromagnetic layer which
can be switched as a whole by the magnetic field generated by the
current passing through the ``write'' lines. Readout is achieved by
measuring the tunnel current through the storage element, which
depends on the magnetization state.  The main difficulty in designing
an MRAM cell is to be able to reliably write bits of information to a
specified cell. One important issue is the problem of 1/2-bit select
due to the close proximity of the ``write'' lines to the free magnetic
layer in the cross-point addressing scheme \cite{akerman05}. Another
issue is the need to use rather strong currents to generate the
magnetic fields required for switching, which makes these devices
quite power-hungry.

Recently, a growing interest was attracted by a new MRAM cell design
concept in which the storage element is made in the shape of a
submicron-sized
ring\cite{zhu00,li01,zhu03,klaui03,castano03,castano04,%
  hayward06,ross06,moneck06,zhu06,vaz07,han08}. As was proposed in
Ref. \onlinecite{zhu00}, the bit may be encoded by the direction of
the vortex magnetization state in a ferromagnetic thin film ring. A
difficulty associated with this design has to do with switching
between the two vortex states, since it requires creating
magnetization poles at the ring boundaries. Hence a rather elaborate
writing scheme using paired word lines\cite{zhu00} or the use of
spin-polarized current\cite{tsoi98,myers99,bass04,klaui05,vaz07} were
called for. Note that in the latter case the current densities
required to affect the magnetization state are found to be quite high
(exceeding $10^8$ A/cm$^2$)\cite{klaui05,vaz07,myers99}. Another
proposal\cite{zhu03,moneck06} has been to pass a current of variable
polarity vertically through the sandwich structure, so that the
resulting circular magnetic field favors a particular vortex
orientation\footnote{Although this design claims substantially lower
  switching currents than those used in Ref. \onlinecite{klaui05}, it
  is doubtful\cite{hubert,vandenberg86,desimone02,moser04} that the
  states with head-on walls and magnetization not aligned with the
  ring boundaries can be feasible in a very soft material like
  permalloy (with the quality factor $Q \simeq 10^{-4}$) used in the
  simulations of Ref. \onlinecite{zhu03}.}.

Here, we propose a new principle for the design of a robust MRAM cell
based on thin film ferromagnetic nanorings (see Fig. \ref{fig:design}
for schematics). In contrast to previous designs, we propose to use
the polarity of the stable $360^\circ$ domain wall in the free
ferromagnetic layer to encode the bit of information. Existence and
stability of such domain walls for certain ranges of parameters was
recently demonstrated by us via micromagnetic modeling and
simulations\cite{mo:jap08}. Winding domain walls, including
$360^\circ$ walls, are also frequently observed experimentally in
cobalt nanorings\cite{ross06,castano03,castano04}.

\begin{figure}
  \centering
  \includegraphics[width=3in]{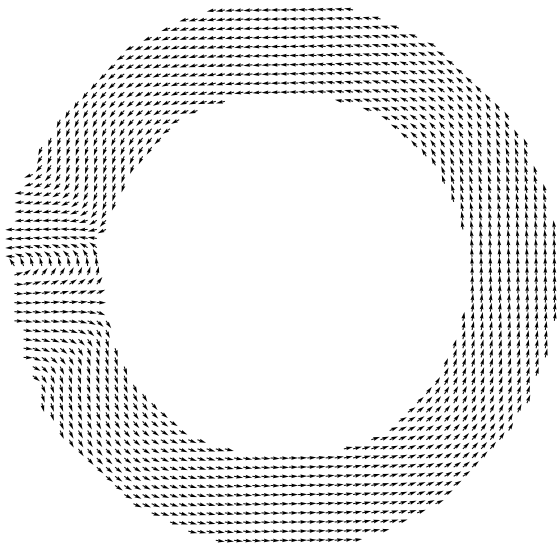}
  \caption{Bit-encoding magnetization state with a $-360^\circ$ wall
    on the left. The other bit is encoded by the mirror image of this
    magnetization pattern, with the $+360^\circ$ wall on the right,
    respectively. From the numerical solution of (\ref{eq:theta}), see
    the text for details.}
  \label{fig:bit}
\end{figure}

The basic idea of the proposed design is that in the presence of a
$360^\circ$ wall the magnetization pattern looks like a vortex state
in most of the ring, except a small region where the wall is tightly
localized (see Fig. \ref{fig:bit}). Nevertheless, in contrast to the
vortex states, its topological degree (i.e. the winding number of the
magnetization vector along a closed loop traced counter-clockwise
inside the ring) is zero, since inside the wall the magnetization
rotates in the direction opposite to the direction of rotation in the
rest of the ring (for vortex states the topological degree is
$+1$). Thus, both states corresponding to the $\pm 360^\circ$ walls
are accessible by smooth in-plane rotation of the magnetization vector
from the ``onion'' state\cite{li01,vaz07,klaui03,dennis02}, the native
magnetization state in the ring following saturation by an in-plane
field. Switching between these states is achieved by passing a current
of variable polarity through a wire running perpendicularly to the
ring plane through its center, generating a circular magnetic field in
the ring, and readout can be performed, as in previous designs, by
arranging the ferromagnetic nanoring to be part of a magnetic tunnel
junction sandwich, with fixed ferromagnetic ring layer in a vortex
state. Note that this design is free from the 1/2-select problem, as
was already pointed out in Ref. \onlinecite{zhu03}.

Let us demonstrate the feasibility of the proposed design by a
micromagnetic study of the Landau-Lifshits-Gilbert
equation\cite{hubert} for the magnetization vector $\mathbf M =
\mathbf M(\mathbf r, t)$ in the ferromagnetic ring (denoted by $\Omega
\subset \mathbb R^3$):
\begin{eqnarray}
  \label{eq:llg}
  {\partial \mathbf{M} \over \partial t} = -{g |e| \over
    2 m c} \left( 
    \mathbf{M} \times \mathbf{H}_\mathrm{eff} + {\alpha \over M_s}
    \mathbf{M} \times \mathbf{M} \times \mathbf{H}_\mathrm{eff} \right),
\end{eqnarray}
with Neumann boundary condition on the material boundary $\partial
\Omega$. The effective field $\mathbf H_\mathrm{eff} = -{\delta E
  \over \delta \mathbf{M}}$, with
\begin{eqnarray}
  \label{eq:Heff}
  && E[\mathbf M] = 
  \int_\Omega \biggl( {K \over 2 M_s^4} M_1^2 M_2^2 + {A   \over 2 
    M_s^2} |\nabla \mathbf M|^2 - \mathbf H \cdot \mathbf M \biggr)
  d^3 \mathbf r \nonumber  \\ && \hspace{0.5cm} +   {1 \over 2}
  \int_{\mathbb 
    R^3}   \int_{\mathbb R^3}   {\nabla   \cdot \mathbf 
    M(\mathbf r) \, \nabla   \cdot \mathbf M(\mathbf r')  \over |\mathbf
    r - \mathbf 
    r'|} \, d^3 \mathbf r \, d^3 \mathbf r'. \label{E}
\end{eqnarray}
Here we assumed the four-fold anisotropy typical of epitaxial films,
e.g., cobalt films\cite{heinrich93}. When the film thickness $d$ is
sufficiently small, this equation can be
reduced\cite{mo:jap08,mo:jcp06} to an effective equation for the angle
$\theta$ between the magnetization vector in the film plane and an
easy direction:
\begin{eqnarray}
  \label{eq:theta}
 \theta_t & = & \Delta \theta - \tfrac14
  \sin 4 
  \theta - h_1 \cos \theta - h_2 \sin \theta \nonumber \\ && + \nu
  \varphi_x  \cos \theta  +  \nu  \varphi_y \sin \theta,
  \\  \label{eq:phi}  \varphi & = & \tfrac{1}{2}
  (-\Delta)^{-1/2} 
  \left(  \theta_x \cos \theta + 
    \theta_y \sin \theta \right)
  \nonumber \\  && + 
  \mathrm{boundary~terms},
\end{eqnarray}
where lengths were rescaled by $L = (A/K)^{1/2}$, and the thin film
parameter\cite{mo:jcp06} $\nu = d /(l Q^{1/2})$ was introduced, where
$l = (A/4 \pi M_s^2)^{1/2}$ is the exchange length and $Q = K /(4 \pi
M_s^2)$ is the material's quality factor.

\begin{figure*}
  \centering
  \includegraphics[width=7in]{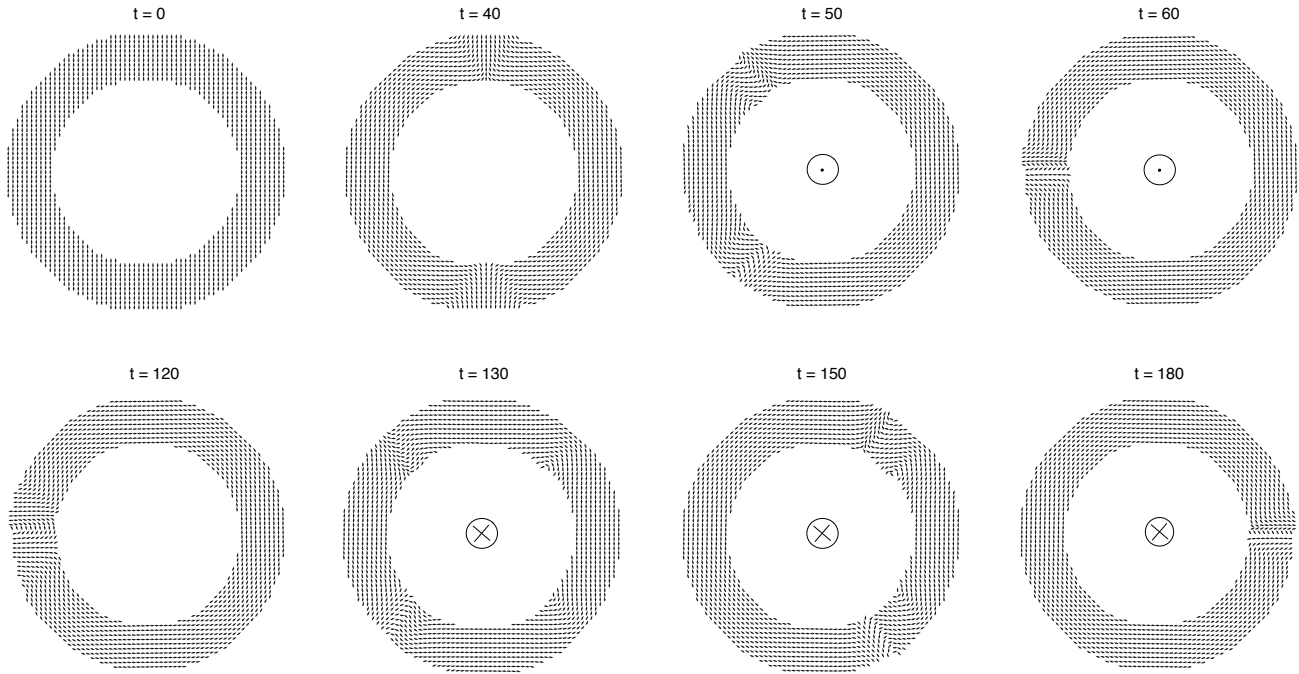}
  \caption{The dynamics of the magnetization reversal in a
    ferromagnetic nanoring subject to the magnetic field from a wire
    passing through the ring center. Results of the numerical
    simulations of (\ref{eq:theta}) (see text for details).}
  \label{fig:ring10}
\end{figure*}

We now investigate the process of switching in a ferromagnetic
nanoring by performing a numerical simulation of (\ref{eq:theta}) in a
ring of diameter 30 and width 5, with $\nu = 10$, and $h_1 = \mp 16 y
/ (x^2 + y^2)$, $h_2 = \pm 16 x / (x^2 + y^2)$, or $h_{1,2} = 0$,
depending on the current through the central wire (all quantities are
dimensionless, see the following paragraph for a physical
interpretation; the details of the numerical method are as in
Ref. \onlinecite{mo:jap08}). At $t = 0$, the magnetization is
saturated by a field along the $y$-axis. As the field is removed, the
magnetization settles to an equilibrium in the form of an onion state
(Fig. \ref{fig:ring10}, $t = 40$). Then, a positive current is applied
through the wire, generating a counter-clockwise magnetic field, which
drives the poles of the onion and the associated charged $180^\circ$
walls towards each other on the left side of the ring. At $t = 60$
these two walls collide to form a $360^\circ$ wall. Importantly, when
the current is stopped, the resulting $360^\circ$ wall maintains its
integrity and does not break up (Fig. \ref{fig:ring10}, $t = 120$,
same as Fig. \ref{fig:bit}, consistently with our earlier studies of
$360^\circ$ walls \cite{mo:jap08}). Applying a negative current leads,
in turn, to a breakup of the $360^\circ$ wall. The resulting charged
$180^\circ$ walls propagate along the ring to collide at the opposite
extreme on the right and form a new $360^\circ$ wall of opposite
polarity (Fig. \ref{fig:ring10}, $t = 180$). As before, when the
current is switched off, the wall remains in its place (not
shown). Note that no magnetic poles need to be created during
switching, the existing poles simply move back and forth along the
ring boundaries.

In conclusion, we have demonstrated the feasibility of storage and
writing of a bit in the form of $\pm 360^\circ$ walls in a thin film
ferromagnetic ring. The dimensionless parameters used in the
simulation can be translated, e.g., to those of a cobalt\cite{li01}
alloy ($l = 3.37$ nm, $M_s = 1400$ emu/cm$^3$, $Q = 0.03$, $d = 5.8$
nm) ring of diameter 585 nm and width 100 nm. The current producing
the required circular magnetic field is $J = 80$ mA, corresponding to
a moderate current density of $7 \times 10^7$ A/cm$^2$. Note that we
used a rather strong circular magnetic field for switching to
accelerate the simulations, considerably weaker fields (and hence
smaller currents) could be used, resulting in a slower switching
process. Another way to reduce the switching current density is to use
a softer material (with smaller $Q$ and $d$), at the expense of
increasing the ring diameter. We emphasize that this design is
expected to be very robust, in view of the stability of the underlying
$360^\circ$ walls\cite{mo:jap08} and the topologically constrained
switching process. Finally, note that the proposed cell geometry
allows to stack the rings vertically by interrupting the vertical wire
with horizontal lines of alternating directions, suggesting a
possibility to greatly increase the memory capacity by harnessing the
third dimension.

The work of C. B. M. was supported, in part, by NSF via grant
DMS-0718027.


\bibliography{../nonlin,../mura,../stat}

\end{document}